\documentclass[12pt]{spieman}  
\usepackage{amsmath,amsfonts,amssymb,mathtools}
\usepackage{graphicx}
\usepackage{setspace}
\usepackage{tocloft}

\title{Microelectromechanical deformable mirror development for high-contrast imaging, part 2: the impact of quantization errors on coronagraph image contrast} 
\usepackage{array}
\newcolumntype{P}[1]{>{\centering\arraybackslash}p{#1}}

\author[a,*]{Garreth~Ruane}
\author[b]{Daniel~Echeverri}
\author[a]{Eduardo~Bendek}
\author[a]{Brian~D.~Kern}
\author[a]{David Marx}
\author[b,a]{Dimitri~Mawet}
\author[a]{Camilo~Mejia~Prada}
\author[a]{A~J~Eldorado~Riggs}
\author[a]{Byoung-Joon Seo}
\author[a]{Eugene~Serabyn}
\author[a]{Stuart~Shaklan}

\affil[a]{Jet Propulsion Laboratory, California Institute of Technology, 4800 Oak Grove Dr., Pasadena, CA 91109, USA}
\affil[b]{Department of Astronomy, California Institute of Technology, 1200 E. California Blvd., Pasadena, CA 91125, USA}

\cftpagenumbersoff{figure}
\cftpagenumbersoff{table}

\graphicspath{{Figures/}}
 
\begin{document} 
  \maketitle 

\begin{abstract}
Stellar coronagraphs rely on deformable mirrors (DMs) to correct wavefront errors and create high contrast images. Imperfect control of the DM limits the achievable contrast and, therefore, the DM control electronics must provide fine surface height resolution and low noise. 
Here, we study the impact of quantization errors due to the DM electronics on the image contrast using experimental data from the High Contrast Imaging Testbed (HCIT) facility at NASA's Jet Propulsion Laboratory (JPL). We find that the simplest analytical model gives optimistic predictions compared to real cases, with contrast up to 3 times better, which leads to DM surface height resolution requirements that are incorrectly relaxed by 70\%. We show that taking into account the DM actuator shape, or influence function, improves the analytical predictions. However, we also find that end-to-end numerical simulations of the wavefront sensing and control process provide the most accurate predictions and recommend such an approach for setting robust requirements on the DM control electronics. From our experimental and numerical results, we conclude that a surface height resolution of approximately 6~pm is required for imaging temperate terrestrial exoplanets around Solar-type stars at wavelengths as small as 450~nm with coronagraph instruments on future space telescopes. Finally, we list the recognizable characteristics of quantization errors that may help determine if they are a limiting factor.
\end{abstract}


{\noindent \footnotesize\textbf{*}Address all correspondence to Garreth Ruane, Email: \linkable{garreth.ruane@jpl.nasa.gov}}\\

\section{Introduction}
\label{sec:intro}  

The light reflected from an exoplanet orbiting a main sequence star can be 10$^{-5}$ to $<$10$^{-10}$ times as bright as the host star depending on the planet type and orbital configuration. 
For instance, a temperate ($\sim$300~K) planet similar to Earth would orbit approximately 0.1~arcseconds from a solar-type star at 10~parsecs and have a planet-to-star flux ratio of $\sim$10$^{-10}$. Imaging and spectroscopy of such a planet requires an optical system that can suppress the diffracted starlight to a similar intensity at the position of the planet. Stellar coronagraphs on future space telescopes, such as the Habitable Exoplanet Observatory (HabEx)\cite{HabEx_finalReport} and Large Ultra-violet, Optical, Infrared Surveyor (LUVOIR)\cite{LUVOIR_finalReport} mission concepts, will make use of one or more deformable mirrors (DMs) to achieve image contrasts of $\sim$10$^{-10}$ in order to directly detect and characterize terrestrial exoplanets orbiting Sun-like stars \cite{Malbet1995,Borde2006,Trauger2007,Pueyo2009}. 

The first step in the process of achieving high contrast is to flatten the wavefront entering the coronagraph. Even with extremely high surface quality optics, typical coronagraphs have contrasts of 10$^{-6}$ or worse in this initial state. The contrast is improved beyond this level using focal-plane wavefront sensing and control algorithms that estimate the stellar field in the image and determine the DM surface changes needed to cancel it based on a model of the optical system. The result is a localized region of high contrast, or ``dark hole," around the star\cite{Groff2015}. Uncorrected wavefront errors cause unwanted starlight to appear within the dark hole, which negatively impacts the sensitivity to faint exoplanets. 

Several DM technologies are under development for use in space-based coronagraph instruments, the most common of which are electrostrictive devices \cite{Ealey2004,Wirth2013} and microelectromechanical systems (MEMS)\cite{Bifano2011,Morgan2019}. A MEMS DM is a metal-coated thin-film mirror whose shape is controlled by an array of electrostatic actuators. The local surface height of the DM is set by the voltage applied to each actuator. While MEMS DMs have achieved promising experimental results, including contrasts of $5\times10^{-9}$ with stability on the order of $10^{-12}$ per hour\cite{MejiaPrada2019}, a few key challenges remain on the path towards readily achieving the 10$^{-10}$ contrast requirement for imaging Earth analogs. Among the critical specifications for all DM technologies, the electronics used to drive the DM surface must be very low noise and allow the DM surface to move in small, well-controlled increments. 

Here, we study the impact of DM quantization errors on the contrast in the image plane. We first develop simple analytical models to predict contrast in an otherwise ideal system. We find that our analytical approach underestimates the impact of quantization errors 
as compared to experimental data from a coronagraph testbed with a MEMS DM in the High Contrast Imaging Testbed (HCIT) facility at NASA's Jet Propulsion Laboratory (JPL). However, higher fidelity numerical simulations of the wavefront sensing and control process are in good agreement with our experimental results. Similar simulations may be used to set robust requirements for the DM surface height resolution for future coronagraph instruments. For instance, we show that a realistic error budget for achieving contrast on the order of 10$^{-10}$ in practice requires $<$10~pm motions per logical bit. In addition, we discuss the characteristics of DM quantization errors that may be used to identify them as a limiting factor in practice. 

\begin{figure}[t]
    \centering
    \includegraphics[width=\linewidth]{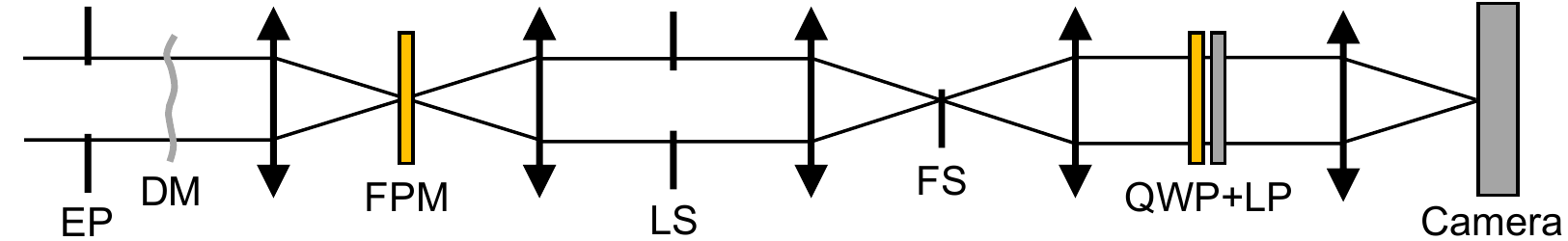}
    \caption{Schematic of the optical system including an entrance pupil (EP) mask, deformable mirror (DM), focal plane mask (FPM), Lyot stop (LS), field stop (FS), quarter-wave plate (QWP), linear polarizer (LP), and imaging camera. The EP mask, DM, LS, QWP, and LP are in collimated space, whereas the FPM and FS are located in focal planes.
    }
    \label{fig:optical_schematic}
\end{figure}

\section{Theory} \label{sec:theory} 

Figure~\ref{fig:optical_schematic} shows a schematic of a coronagraph instrument, which consists of a DM, a focal plane mask (FPM), and a Lyot stop (LS) in a subsequent pupil plane. The DM is used to create a dark hole in an image plane after the LS. The light outside of the dark hole may be blocked using a field stop (FS) such that the optics downstream of the FS, which often includes polarizers and spectral filters, have relaxed manufacturing requirements. The plane of the FS is re-imaged onto the camera to create the final high-contrast image. 

The coronagraph may be described as a linear operator that propagates the field in the pupil-plane containing the DM, $E_p(x,y)$, to the final image: $E_f(\xi,\eta) = C\{E_\text{p}(x,y)\}$, where $(x,y)$ and $(\xi,\eta)$ are the pupil and image plane coordinates, respectively. The raw contrast at a given position is defined as the ratio between the intensity due to an on-axis source (the star) and an equivalent source at that position\cite{Ruane2018_metrics}. The normalized intensity, $\hat{I}(\xi,\eta)$, is divided by the intensity of an off-axis source at a representative position in the image. In the following, we assume that the DM is nominally in the state that creates a dark hole in the image, which minimizes the normalized intensity.

A simple and widely-used analytical model for the stellar intensity uses a modal argument\cite{TraubOppenheimer2010}. With slight modification to the notation, the normalized intensity is estimated by
\begin{equation}
    \hat{I} = \pi\left(  \frac{8h_\text{rms}}{n_\text{act}\lambda} \right)^2,
    \label{eqn:TraubOppInorm}
\end{equation}
where $h_\text{rms}$ is the RMS surface error, $n_\text{act}$ is the number of actuators across the LS, and $\lambda$ is the wavelength. Uniformly distributed quantization errors are expected to have $h_\text{rms} = h_\text{min}/\sqrt{12}$, where $h_\text{min}$ is the minimum DM surface motion enabled, and thus we write this model as 
\begin{equation}
    \hat{I} = \frac{16\pi}{3 n_\text{act}^2} \left(\frac{h_\text{min}}{\lambda} \right)^2
    \label{eqn:TraubOppInorm2}
\end{equation}
or more simply as $\hat{I}=(h_\text{min}/h_0)^2$, where
\begin{equation}
    h_0 = \sqrt{\frac{3}{16\pi}}n_\text{act}\lambda.
    \label{eqn:h0TraubOpp}
\end{equation}
However, this expression ignores the DM actuator shape, or influence function, which has a significant impact on the distribution of the stellar intensity in the image. 

\begin{figure}[t]
    \centering
    \includegraphics[width=\linewidth]{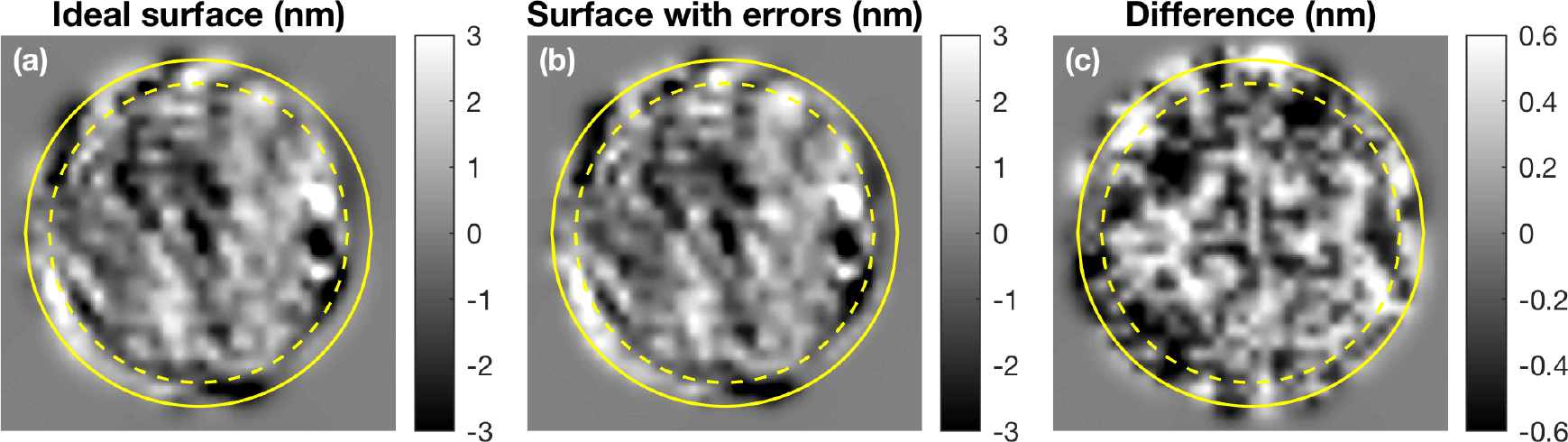}
    \caption{Example deformable mirror (DM) surfaces. (a)~Desired DM surface. (b)~DM surface with quantization errors due to a minimum motion of 1~nm. (c)~The difference between (a) and (b). The solid and dashed circles indicate the relative size of the EP mask and LS, respectively, in both our simulations and testbed.}
    \label{fig:DMshapes}
\end{figure}

To improve upon this model, we derived an alternate expression that treats the DM surface as a linear superposition of actuators with influence function $f_\text{infl}(x,y)$:
\begin{equation}
    h(x,y) = \sum_{j=1}^{N_\text{act}} V_j \;g_j\;f_\text{infl}(x-x_j,y-y_j),
\end{equation}
where $N_\text{act}$ is the total number of actuators, $V_j$ are the voltage settings, $g_j$ are the gain coefficients (i.e. the surface displacement per volt), and $(x_j,y_j)$ are the coordinates of the actuator centers. The least significant bit (LSB) provided by digital-to-analog converter in the DM electronics limits the smallest amount a single actuator can be moved to $h_\text{min}=g_j V_\text{LSB}$, where $V_\text{LSB}$ is the voltage difference of the LSB. We do not consider methods that use high-frequency bit-cycling to improve upon the actuator surface height resolution. Figure \ref{fig:DMshapes} shows an example of a DM surface with quantization errors corresponding to a minimum actuator motion of 1~nm. 

For simplicity, we assume that $h_\text{min}$ is uniform for all actuators and the influence function is a peak-normalized Gaussian $f_\text{infl}(x,y) = \exp(-(r/d)^2)$, where $r^2 = x^2 + y^2$. The radius of actuator influence, $d$, is related to the inter-actuator pitch, $p$, by the parameter $\omega=d/p$, which depends on the DM architecture and may need to be determined empirically. In appendix~\ref{sec:appenA}, we derive the following expression for the normalized intensity due to quantization errors:
\begin{equation}
    \hat{I}(\alpha) = \frac{16\pi}{3n_\text{act}^2} \left(\frac{h_\text{min}}{\lambda} \right)^2 \pi^2\omega^4  \exp\left(-\left(\alpha/\alpha_\text{infl}\right)^2\right),
    \label{eqn:inorm}
\end{equation}
where $\alpha$ is the angular separation from the optical axis 
and 
\begin{equation}
    \alpha_\text{infl} = \frac{\sqrt{2}}{2\pi}\frac{\lambda}{d} = \frac{\sqrt{2}}{2\pi}\frac{n_\text{act}}{\omega}\frac{\lambda}{D_\text{LS}} = \frac{\sqrt{2}}{2\pi}\frac{n_\text{act}}{\omega\Gamma}\frac{\lambda}{D},
    \label{eqn:alphainfl}
\end{equation}
where $D_\text{LS}$ is the diameter of the Lyot stop, $D$ is the full pupil diameter, and $\Gamma=D_\text{LS}/D$. The full width at half maximum of the influence function and the transfer function (i.e. the Fourier transform of the influence function) are 1.67$d$ and 1.67$\alpha_\text{infl}$, respectively. 

There are two differences between Eqn.~\ref{eqn:inorm} and the simpler expression in Eqn.~\ref{eqn:TraubOppInorm2}. First, the normalized intensity is scaled by $\pi^2\omega^4$. Secondly, Eqn.~\ref{eqn:inorm} includes the intensity fall off due to the actuator transfer function. In fact, Eqn.~\ref{eqn:inorm} simplifies to  Eqn.~\ref{eqn:TraubOppInorm2} if $\omega=1/\sqrt{\pi}$ and $\alpha\approx0$. In appendix~\ref{sec:appenA}, we also present an extension to Eqn.~\ref{eqn:inorm} that models influence function as the sum of two Gaussians in order to investigate the impact of the influence function shape (see Eqn. \ref{eqn:Itwogaussians}). In the following sections, we compare each of these expressions to experimental measurements. 

\section{Experimental Method}\label{sec:method} 

To demonstrate the impact of the DM quantization errors experimentally, we used a coronagraph testbed whose primary purpose was to test vortex coronagraphs\cite{Mawet2005,Foo2005,Ruane2018_JATIS,SerabynTDEM2} at high contrast. For this work, we artificially injected quantization errors by rounding the DM commands to move each DM actuator by a discrete amount representing a uniform $h_\text{min}$. We then varied the effective $h_\text{min}$ to study its impact on the normalized intensity in the dark hole. 

We used the open-source and freely available Fast Linearized Coronagraph Optimizer (FALCO) toolbox 
for the wavefront sensing and control on the testbed\cite{Riggs2018}. FALCO's wavefront control loop uses pair-wise probing\cite{Giveon2011} to estimate the electric field in the image plane and electric field conjugation (EFC)\cite{Giveon2009} to determine the DM settings that cancel the field. Since this is a model-based approach, FALCO makes use of a wave propagation model that calculates the electric field in the image plane for a given DM setting taking into account the coronagraph masks. This model can also be used separately to simulate the testbed in a standalone fashion. 

The experimental setup consisted of a supercontinuum laser source that was circularly polarized and focused onto a pinhole to create a simulated star. The light from the pinhole was collimated by an off-axis parabolic mirror (OAP). The remainder of the optical system is illustrated in Fig.~\ref{fig:optical_schematic}. A circular entrance pupil (EP) mask defined the pupil 63~mm upstream of a Boston Micromachines Kilo-DM with 952 actuators with inter-actuator spacing of $p=300~\mu$m. The DM was controlled using 16-bit electronics manufactured by Teilch set to provide a range of 100~V\cite{Bendek2020}. The circular beam illuminated an area that was 29.8 actuators across on the DM. The focal plane mask (FPM) was a charge-4, liquid-crystal, vector vortex waveplate\cite{Serabyn2019}. The radius of the LS was $\Gamma$~=~86.3\% of the full geometric image of EP and thus there were effectively $n_\text{act}=25.7$ actuators across the LS. 
We measured the DM actuator gains using a Fizeau interferometer (Zygo Verifire) by poking isolated actuators with the DM in its flat state, which is a nominal DM setting ($\sim$80~V peak-to-valley) that removes low order aberrations that appear when the DM is unpowered. Since the surface deflection is quadratic with voltage, the actuator gains are nonuniform and range from 2 to 4 nm/V across the illuminated region with an average of $\sim$3.3 nm/V. The FS was a razor blade edge that blocked more than half of the image plane including the bright central core at the position of the pseudo star and the dark hole was created on the transmitted side. After the FS, the light passes through a quarter-wave plate (QWP) and linear polarizer (LP) to filter an unwanted circular polarization that was not diffracted by the FPM. All of the powered optics were reflective OAPs to minimize chromatic aberrations. The optical table was inside of a vacuum chamber at $\sim$100~Torr.  

Starting with the DM in its flat state, we ran 10 iterations of the EFC algorithm in a single spectral band (20~nm bandwidth centered at 670~nm), which is sufficient to converge to a normalized intensity that is dominated by an incoherent background at a normalized intensity of $\sim10^{-8}$. We then repeated this process with artificial DM quantization errors ranging from 0.1~nm to 2.0~nm in steps of 0.1~nm. The DM quantization errors were injected by rounding the DM commands such that individual actuators can only move in steps of $h_\text{min}$. The dark hole was generated over 2-12~$\lambda/D$ in a partial annulus with a 140$^\circ$ opening angle.

\begin{figure}[t]
    \centering
    \includegraphics[width=\linewidth]{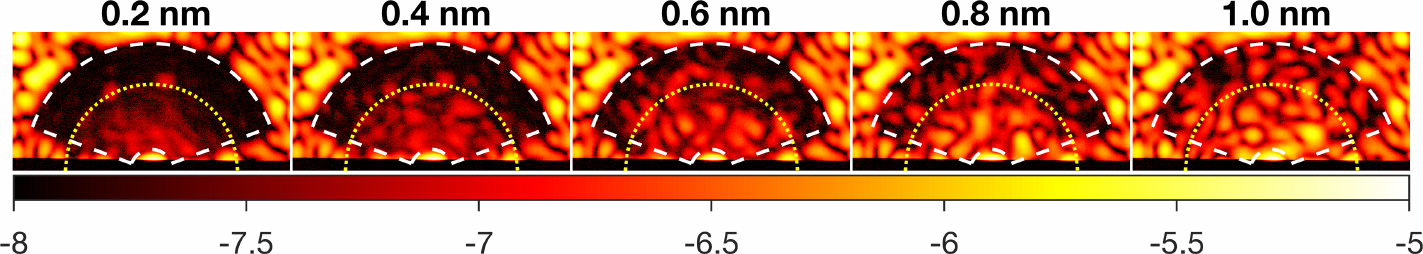}
    \caption{Experimental measurements of the log normalized intensity in the dark hole region versus the minimum DM surface motion allowed, $h_\text{min}$. The dashed white lines indicate the intended dark hole, which is a partial annulus with and inner and outer radii corresponding to 2 and 12 $\lambda/D$ from the star, respectively, and and opening angle of 140$^\circ$. The dotted yellow semi-circle shows the radius of $\alpha_\text{infl}=8.15~\lambda/D$ (see Eqn.~\ref{eqn:alphainfl}). The horizontal black stripe across the bottom of the images is the shadow of the field stop (FS in Fig.~\ref{fig:optical_schematic}).}
    \label{fig:dhs_vs_hmin}
\end{figure}

\section{Results}\label{sec:results} 

\begin{figure}[t]
    \centering
    \includegraphics[width=0.5\linewidth]{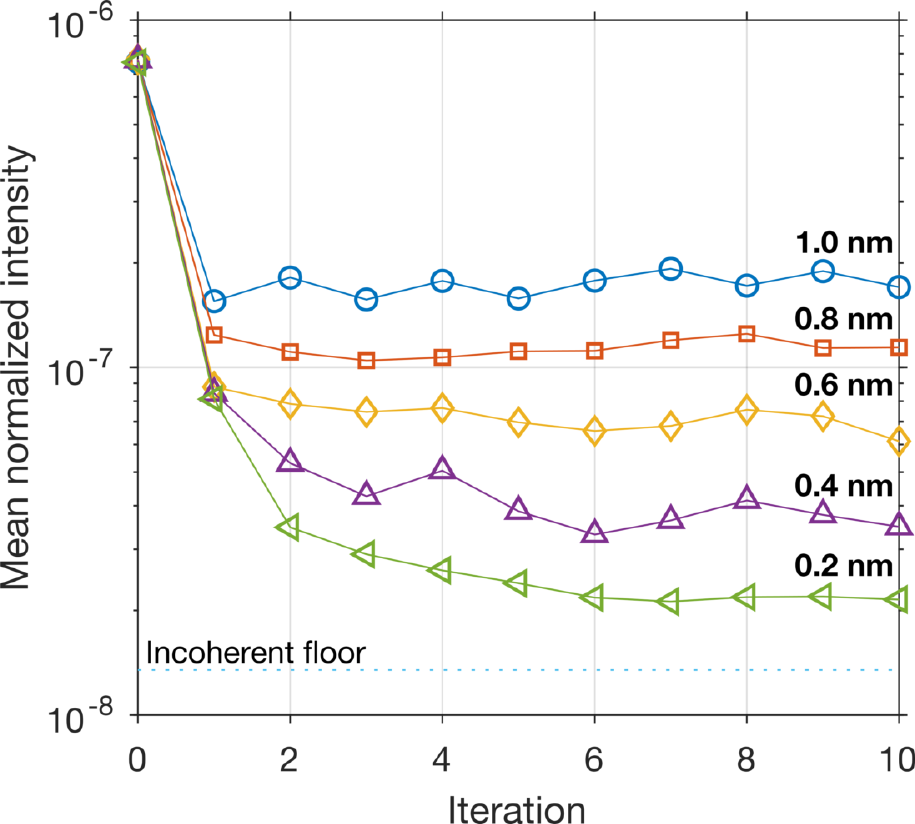}
    \caption{Experimental measurements of the normalized intensity versus EFC iteration after artificially injecting quantization noise with minimum DM motion ranging from 0.2-1.0~nm. The dotted line shows the level of the incoherent light limiting the testbed contrast.}
    \label{fig:normI_vs_it_tb_examples}
\end{figure}

Figure \ref{fig:dhs_vs_hmin} shows the normalized intensity measured on the testbed for representative $h_\text{min}$ values. The quantization errors introduce speckles in the dark hole. To ensure that these speckles are indeed due to the injected errors, we ran more iterations than necessary for reaching the contrast floor. Figure~\ref{fig:normI_vs_it_tb_examples} shows the mean normalized intensity in the dark hole as a function of EFC iteration for the cases shown in Fig.~\ref{fig:dhs_vs_hmin}. While we ran 10 iterations by default, the algorithm converged in approximately five iterations in all cases. During these experiments, temporally incoherent light appeared on the testbed just above the 10$^{-8}$ level, which is independent of the errors we injected. 

\begin{figure}[t]
    \centering
    \includegraphics[width=0.5\linewidth]{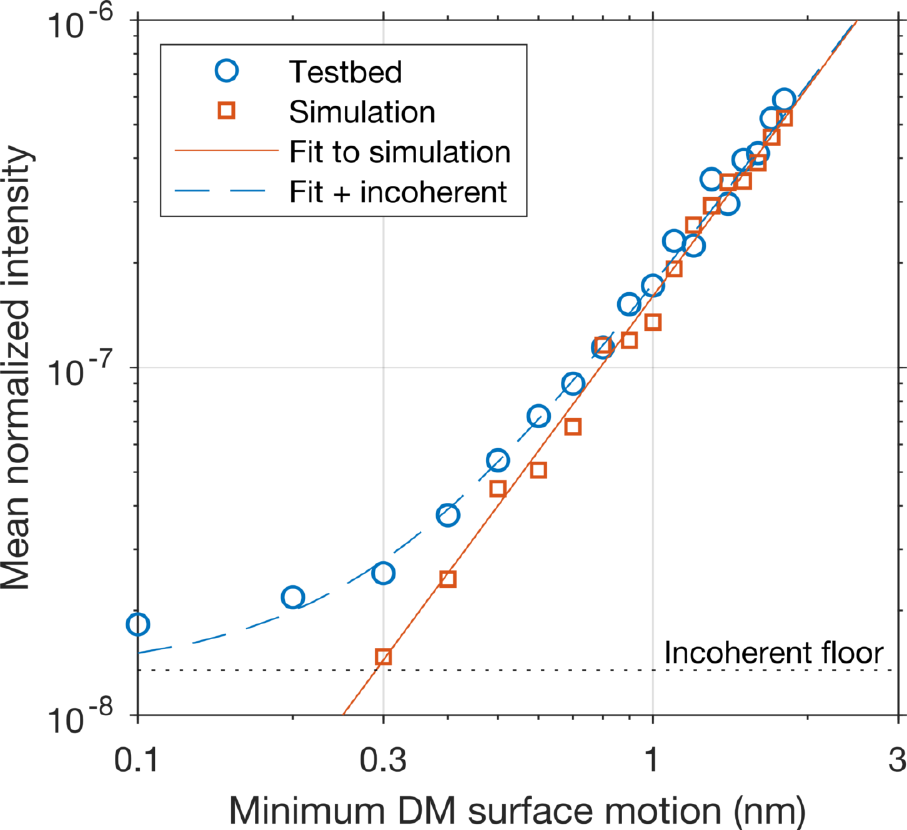}
    \caption{Comparison between normalized intensity measured on the testbed (circles) versus the full numerical propagation simulation of the optical system (squares). The solid line is the best fit parabola to the simulation results. The dashed line shows the fit added to the incoherent floor measured on the testbed. }
    \label{fig:normI_vs_hmin}
\end{figure}

Figure~\ref{fig:normI_vs_hmin} shows the mean normalized intensity in the dark hole versus $h_\text{min}$ for both the testbed and the corresponding numerical simulation using the testbed model in FALCO. For each value of $h_\text{min}$, we took the median of the last three EFC iterations and fit a second order power law of the form $(h_\text{min}/h_0)^2$ in each case. This resulted in $h_0 = 2390\pm60$~nm for the testbed data and $h_0 = 2500 \pm 40$~nm using the FALCO simulation, where the error bars are the 95\% confidence bounds from the fit. We ignored the first five data points for the testbed case (i.e. cases with $h_\text{min}<$~0.5~nm) to prevent the incoherent component from biasing the fit results. For comparison, adding the incoherent component to the simulation fit leads to good agreement with the testbed results (see dashed line in Fig.~\ref{fig:normI_vs_hmin}).

\section{Discussion}\label{sec:discussion} 

\subsection{Experimental results vs. analytical predictions}

There is a minor discrepancy between the $h_0$ values derived from fitting the testbed data and the FALCO simulation. We attribute the difference to imperfect calibration of the DM actuator gains in the model. This calibration error mainly results in a small impact on the convergence rate of the EFC algorithm. Otherwise, this difference is not significant for the purposes of the following discussion. 
There is a much larger discrepancy between the results above and analytical models. 
For instance, compared to Eqn.~\ref{eqn:TraubOppInorm}, with $n_\text{act}$~=~25.7 and $\lambda$~=~670~nm, we find $h_0$~=~4206~nm, which means the predicted normalized intensity is $(4206/2390)^2$~=~3.1$\times$ smaller than the testbed. 

\begin{figure}[t]
    \centering
    \includegraphics[width=0.5\linewidth]{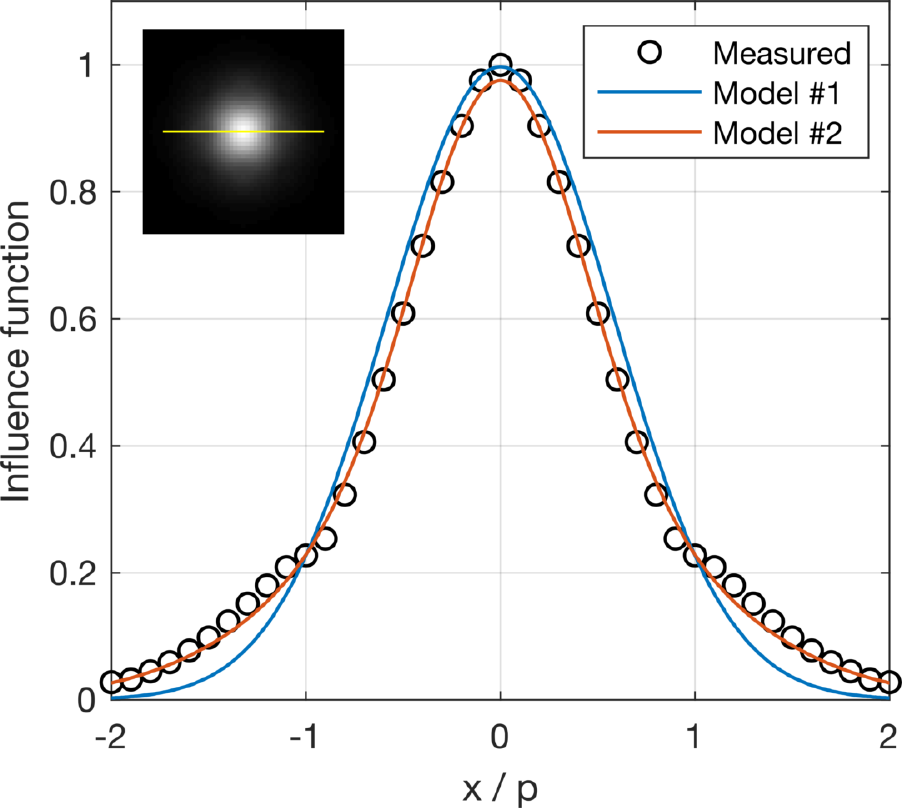}
    \caption{The DM influence function and best fit models using a Gaussian (Model~\#1) and a sum of two Gaussians (Model~\#2). The inset shows the full 2D influence function over a 5$\times$5 actuator region. The yellow line indicates the direction of the line profile, which is along the actuator grid. }
    \label{fig:infl_func}
\end{figure}

Equation~\ref{eqn:inorm} is based on a model of the DM that includes the influence function characteristic width, which is especially important for MEMS DMs with a continuous surface because the influence function width is significantly larger than the actuator pitch.  This is the reason the peak-to-valley of the DM surface errors in Fig.~\ref{fig:DMshapes}c exceeds the corresponding $h_\text{min}$ value of 1~nm. Figure~\ref{fig:infl_func} shows the best fit Gaussian profile to the measured DM influence function in the least-squares sense, which gives $\omega=d/p=0.82$ and a root-mean-square error (RMSE) of 2.4\% computed over 6.6$\times$6.6 actuator region. In order to compare Eqn.~\ref{eqn:inorm} with our experimental results and other analytical models, we take the average value in the dark hole region; the term $\exp\left(-\left(\alpha/\alpha_\text{infl}\right)^2\right)$ averages to $\phi=0.39$ over 2-12~$\lambda/D$ with $\alpha_\text{infl}=8.15~\lambda/D$. Thus, Eqn.~\ref{eqn:inorm} may be written in the form of $(h_\text{min}/h_0)^2$ with
\begin{equation}
    h_0 = \sqrt{\frac{3}{16\pi}}\frac{n_\text{act}\lambda}{\pi\omega^2\sqrt{\phi}},
    \label{eqn:h0gauss}
\end{equation}
which is equivalent to Eqn.~\ref{eqn:h0TraubOpp} divided by $\pi\omega^2\sqrt{\phi}=1.33$. In this case, our analytical result predicts $h_0$~=~3160~nm and the dark hole intensity is a factor of $(3160/2390)^2$~=~1.7 times lower than our experimental measurements, but a factor of $\pi^2\omega^4\phi=1.8$ higher than Eqn.~\ref{eqn:TraubOppInorm}. 

One potential shortcoming of Eqn.~\ref{eqn:h0gauss} is that it is based on a relatively poor model of the actuator influence function. In appendix~\ref{sec:appenA}, we also show that our analytical expression may be easily generalized to model the influence function as the sum of two Gaussian functions, which gives a better fit to the measured influence function (RMSE of 0.47\% versus 2.4\%). The best fit parameters are $c_1=0.60$, $c_2=0.38$, $\omega_1=0.59$, and $\omega_2=1.23$ (see Fig.~\ref{fig:infl_func}). Again, to cast the normalized intensity in the form $(h_\text{min}/h_0)^2$:
\begin{equation}
    h_0 = \sqrt{\frac{3}{16\pi}}\frac{n_\text{act}\lambda}{\pi \sqrt{\phi^\prime}},
    \label{eqn:h0gauss2}
\end{equation}
where $\phi^\prime=0.16$ is the spatial mean of $\Phi(\alpha)$ over 2-12~$\lambda/D$ (see appendix~\ref{sec:appenA} for definition of $\Phi$). For comparison, the equivalent $\phi^\prime$ parameter in the single Gaussian case is $\phi^\prime=\omega^4\phi$~=~0.18. This higher fidelity model actually predicts a smaller dark hole intensity than the single Gaussian case with $h_0$~=~3366~nm and an intensity that is $(3366/2390)^2$~=~2 times smaller than our experimental measurements. Table~\ref{tab:compare_analyt} summarizes the results from the testbed, numerical simulation, and analytical models.

\begin{table}[h!]
\caption{Comparison between our results and analytical models in terms of the predicted $h_0$ parameter, normalized intensity for a representative error of $h_\text{min}$~=~0.1~nm calculated via $(h_\text{min}/h_0)^2$, and the maximum $h_\text{min}$ for achieving $\hat{I}$~=~10$^{-10}$ assuming an otherwise perfect system.}
\label{tab:compare_analyt}
\begin{center}       
\begin{tabular}{|c|c|c|c|}
\hline
\rule[-1ex]{0pt}{3.5ex} Method & $h_0$~(nm) & $\hat{I}$ for $h_\text{min}$~=~0.1~nm & $h_\text{min}$ for $\hat{I}$~=~10$^{-10}$ (pm)\\
\hline\hline
\rule[-1ex]{0pt}{3.5ex} Testbed & 2390 & 1.8$\times$10$^{-9}$ & 24 \\
\hline
\rule[-1ex]{0pt}{3.5ex} FALCO simulation & 2500 & 1.6$\times$10$^{-9}$ & 25 \\
\hline
\rule[-1ex]{0pt}{3.5ex} Traub \& Oppenheimer (2010) & 4206 & 5.7$\times$10$^{-10}$ & 42 \\
\hline
    \rule[-1ex]{0pt}{3.5ex} Eqn.~\ref{eqn:inorm} (Gaussian) & 3160 & 1.0$\times$10$^{-9}$ & 32 \\
\hline
\rule[-1ex]{0pt}{3.5ex} Eqn.~\ref{eqn:Itwogaussians} (Sum of 2 Gaussians) & 3366 & 8.8$\times$10$^{-10}$ & 34 \\
\hline
\end{tabular}
\end{center}
\end{table} 

\subsection{Other potential shortcomings of the analytical models}

The analytical models above underestimate the impact of quantization errors on the contrast compared to the experimental data and end-to-end numerical simulations. Here, we discuss potential shortcomings of these models that could account for these discrepancies. 

In appendix~\ref{sec:appenA}, we show that the field in the image plane may be written as
\begin{equation}
    E_f(\xi,\eta) = C\left\{E_p^\text{dh}(x,y)\right\} + i2k C\left\{E_p^\text{dh}(x,y)\Delta h(x,y)\right\},
\end{equation}
where $C\{.\}$ is the coronagraph operator, $E_p^\text{dh}(x,y)$ is the field in the input pupil that generates a dark hole in the image plane, $C\left\{E_p^\text{dh}(x,y)\right\}$ is the field in the dark hole without quantization errors, $k=2\pi/\lambda$, and $\Delta h(x,y)$ is the DM surface error. To derive the analytical models presented above, we assume that the first term is negligible when the quantization errors dominate. However, during the EFC process, each iteration only removes a portion of the residual starlight in the dark hole, and in practice $C\left\{E_p^\text{dh}(x,y)\right\}$ never becomes truly negligible with respect to the contribution due to the quantization errors. When the wavefront control algorithm reaches the intensity floor, EFC may only reduce the intensity by a factor of a few at each iteration even with full control of the DM, whereas the second term is approximately constant. If the first term is 2 times smaller than the second, then we would expect the actual stellar intensity to be $(3/2)^2$~=~2.25 times higher than our idealized analytical model, which is plausible given practical EFC convergence rates. 

While the uncorrected stellar field may fully account for the discrepancies above, there are other assumptions that may break down in some cases. For instance, the influence of actuators outside of the Lyot stop may not be negligible in some coronagraphs and some DMs may not be well modeled by a superposition of independent influence functions. Nonetheless, the models can be compared to any high-contrast system by injecting known quantization errors using a similar approach to the experimental method above.

\subsection{Requirements on the DM electronics}

The wavefront error requirements for future space telescopes are typically derived based on the maximum allowable change in raw contrast at any position within the high-contrast field of view (i.e. dark hole)\cite{HabEx_finalReport,LUVOIR_finalReport}. The quantization errors will likely have the most impact near the inner working angle and at the shortest wavelengths\cite{Ruane2018_JATIS}. Since our experimental measurements are based on the mean contrast over the dark hole at $\lambda_0$~=~670~nm, we scale the contrast by $\lambda^{-2}$ to predict the contrast at other wavelengths and estimate the contrast at the inner working angle by assuming a Gaussian influence function (see Eqn.~\ref{eqn:inorm}): 
\begin{equation}
    \hat{I}(\alpha) \approx \left(\frac{h_\text{min}}{h_0^\prime} \right)^2 = \left(\frac{h_\text{min}}{h_0} \right)^2 \left(\frac{\lambda_0}{\lambda} \right)^2 \frac{1}{\phi} \exp\left(-\left(\alpha/\alpha_\text{infl}\right)^2\right),
    \label{eqn:inormapprox}
\end{equation}
where $h_0^\prime$ is the corrected version of the $h_0$ parameter:
\begin{equation}
    h_0^\prime = h_0 \sqrt{\frac{\lambda}{\lambda_0}}\sqrt{\phi}\exp\left(\frac{1}{2}\left(\alpha/\alpha_\text{infl}\right)^2\right).
    \label{eqn:h0prime}
\end{equation}
For instance, the HabEx coronagraph has a shortest wavelength of 450~nm, an inner working angle of approximately 3~$\lambda/D$. By extrapolating our experimental results, which assumes HabEx uses the same DM as our testbed, we find that $h_0^\prime$~=~0.55$h_0$~=~1309~nm. 

To build up a realistic error budget for achieving a raw contrast of 10$^{-10}$, we may choose to allocate a intensity residual of $I_\text{req}$~=~2$\times$10$^{-11}$ to quantization errors, which leads to a requirement of $h_\text{min}=\sqrt{I_\text{req}}h_0^\prime=\sqrt{2\times10^{-11}}\times$1309~nm~=~6~pm. In general, the $h_\text{min}$ requirement scales with the square root of the allocated contrast. Flattening our MEMS DM requires a nominal voltage map that mostly removes the natural defocus shape of the DM surface and ranges from 0~V to 80~V. After running the EFC algorithm to create the dark hole on our testbed, the nominal voltage map changes by approximately $\pm$1~V. The DM electronics allow a full range of 100~V and 2$^{16}$ levels (i.e. 16 bits). With a gain of 4~nm/V (i.e. the worst case gain for our MEMS DM), the theoretical $h_\text{min}$ is 6~pm and our current electronics are in principle capable of sufficient surface height resolution. However, other types of electronic noise may limit our ability to control the surface height to single bit precision. 

While vortex coronagraphs use little stroke to achieve a dark hole by design, coronagraphs that achieve high contrast, such as the Lyot coronagraphs used on HCIT testbeds\cite{DSTroadmap}, can require up to a half-wave of stroke to create a dark hole. Assuming a maximum wavelength of 800~nm, the EFC algorithm could apply up to $\pm$50~V. If the full voltage range is increased to 150~V to accommodate this, the theoretical $h_\text{min}$ increases to 9~pm, which violates the requirement above. There are a few possible solutions to this: (1)~add an additional bit, (2)~improve the natural shape of the mirror to reduce the voltage needed to flatten the DM, or (3)~allow the DM surface to take on its natural shape and compensate for the associated defocus with static optical alignment in order to minimize the nominal voltage otherwise used to completely flatten the DM. 

The error budget presented above assumes the DM is similar to the BMC Kilo DM on our testbed, but larger-format DMs will likely be used for future space telescopes. The requirements for the HabEx coronagraph, for instance, include 64$\times$64 DMs which translates to $n_\text{act}$~=~59 assuming 62 actuators across the entrance pupil and $\Gamma$~=~0.95\cite{Krist2019}. This increases $n_\text{act}$ by a factor of $59/25.7$~=~2.3 and, by Eqn.~\ref{eqn:inorm}, scales the stellar intensity by $1.11(25.7/59)^2$~=~0.21, where the factor of 1.11 is due to the widening of the transfer function profile. HabEx will also use two DMs in series, which effectively doubles the total number of actuators within the Lyot stop, denoted $ N_\text{act}^\prime$ in appendix~\ref{sec:appenA}, and thereby doubles the intensity due to quantization errors. These differences between the HabEx coronagraph and our testbed combine to scale the intensity by a factor of 0.42 and effectively relax the $h_\text{min}$ requirement from 6~pm to 9~pm. Further refinement of this requirement would also need to account for changes in the influence function shape because the $h_\text{min}$ requirement will scale with $\omega^2$. Ultimately, the final requirement should be derived from an end-to-end model, or testbed, with the actual HabEx DM configuration. 

\subsection{Identifying quantization errors in practice}

Quantization errors are extremely small and may be difficult to identify using traditional metrology methods. However, the characteristics of the stellar field in the dark hole can provide strong evidence that the contrast is limited by quantization errors. In such cases, when a coronagraph approaches its best possible contrast:
\begin{enumerate}
    \item 
    The field estimated by pairwise probing\cite{Giveon2011} will account for most of the stellar intensity and will appear randomized at each iteration. 
    \item The normalized intensity in the dark hole will decrease with separation following the transfer function $\exp\left(-\left(\alpha/\alpha_\text{infl}\right)^2\right)$.
    \item The normalized intensity will reduce with wavelength. 
\end{enumerate}
Artificially injecting quantization noise during the wavefront control process either in the real instrument of end-to-end model can be a useful approach for determining the nature of quantization errors for a particular system. 


\section{Conclusion} \label{sec:conc} 

We have experimentally determined the relationship between DM quantization errors and coronagraph image contrast. We showed that end-to-end numerical modeling of the coronagraph optical system, including the DM influence function, provides the best predictions of the contrast performance. On the other hand, the simplest analytical models tend to underestimate the stellar intensity in the dark hole (by up to a factor of 3), but accurately representing the shape of the DM influence function in the analytical models mitigates much of the discrepancy. These errors may have a significant impact on the DM surface height resolution requirements and thus the design of the high-voltage electronics for coronagraph instruments on future space telescopes. We argue that allocating a contrast floor of 2$\times$10$^{-11}$ to quantization errors leads to a surface height resolution requirement of 6~pm, which can be achieved using 16-bit electronics with a range of 100~V and a DM with gain of 4~nm/V. We have also described the characteristics of DM quantization errors in coronagraph images that may help identify them as a limiting factor in future work.


\appendix

\section{Derivation of the analytical model}\label{sec:appenA}



\subsection{The optical system} 

The coronagraph consists of a DM that is conjugate to the telescope pupil, a vortex focal plane mask, and a circular Lyot stop. Since these planes are related by optical Fourier transforms, we define a linear operator that propagates the field in the pupil-plane containing the DM, $E_p(x,y)$, to the final image: 
\begin{equation}
    E_f(\xi,\eta) = C\{E_\text{p}(x,y)\} = \text{FT}\left\{\text{FT}\left\{\text{FT}\left\{E_\text{p}(x,y)\right\}e^{il\theta}\right\}L(x,y)\right\},
\end{equation}
where 
\begin{equation}
    \text{FT}\{E(x,y)\} = \frac{1}{\lambda f}\iint E(x,y) e^{-ik(x\xi+y\eta)/f} dx dy,
\end{equation}
$k=2\pi/\lambda$, $\lambda$ is the wavelength, $f$ is the focal length, and $L(x,y)$ is the Lyot stop function. Here, the Lyot stop is a simple circular aperture that is undersized with respect to the geometric beam.  

\subsection{Propagation of small DM voltage errors} 

We assume the wavefront control algorithm successfully determines the voltage settings to generate a dark hole in the final image plane, but the electronics introduce errors, $\Delta V_j$, about the ideal DM settings. The resulting DM surface may be represented as $ h(x,y) = h_\text{dh}(x,y) + \Delta h(x,y)$, where $h_\text{dh}(x,y)$ is the DM surface that provides a dark hole and $\Delta h(x,y)$ is the surface height error: 
\begin{equation}
    \Delta h(x,y) = \sum_{j=1}^{N_\text{act}}\Delta V_j\;g_j\;f_\text{infl}(x-x_j,y-y_j).
    \label{eqn:deltah}
\end{equation}
The stellar field in the final image plane is given by 
\begin{equation}
    E_f(\xi,\eta) = C\left\{E_p(x,y)e^{i2k h(x,y)}\right\} = C\left\{E_p^\text{dh}(x,y)e^{i2k \Delta h(x,y)}\right\},
\end{equation}
where $E_p^\text{dh}(x,y) = E_p(x,y)\exp(i2k h_\text{dh}(x,y))$. Assuming $2k\Delta h(x,y) \ll 1~\text{rad}$,
\begin{equation}
    e^{i2k \Delta h(x,y)} \approx 1 + i2k \Delta h(x,y)
\end{equation}
and
\begin{equation}
    E_f(\xi,\eta) = C\left\{E_p^\text{dh}(x,y)\right\} + i2k C\left\{E_p^\text{dh}(x,y)\Delta h(x,y)\right\}.
    \label{eqn:ef}
\end{equation}

\subsection{Estimated contrast in an otherwise ideal system}

When the DM quantization error dominates, the first term in Eqn.~\ref{eqn:ef} is negligible and
\begin{equation}
    E_f(\xi,\eta) \approx i2k C\left\{E_p^\text{dh}(x,y)\Delta h(x,y)\right\}.
\end{equation}
In a vortex coronagraph, the pupil field is approximately an evenly illuminated plane wave and the coronagraph only filters out low-order aberrations. The contrast in the dark hole is mostly impacted by mid-spatial frequency aberrations. Assuming $\Delta h(x,y)$ has negligible power at low spatial frequencies (i.e. $\lesssim$2~cycles per pupil diameter), the field is approximately
\begin{equation}
    E_f(\xi,\eta) \approx i2k\;\text{FT}\left\{ L(x,y) \Delta h(x,y) \right\}.
\end{equation}
Combining with Eqn.~\ref{eqn:deltah}, 
\begin{equation}
    E_f(\xi,\eta) = i2k \sum_{j=1}^{N_\text{act}}\Delta V_j\;g_j\;\text{FT}\left\{ L(x,y) f_\text{infl}(x-x_j,y-y_j) \right\}.
\end{equation}
We approximate the impact of the Lyot stop by only summing over the number of actuators whose center is within the Lyot stop opening, $N_\text{act}^\prime$.
In other words, we assume the actuators outside of the Lyot stop have a negligible impact and the field simplifies to
\begin{equation}
    E_f(\xi,\eta) \approx i2k \;\text{FT}\{f_\text{infl}(x,y)\} \sum_{j=1}^{N_\text{act}^\prime}\Delta V_j\;g_j\;e^{-ik(x_j\xi+y_j\eta)/f}.
\end{equation}
Thus, the image plane intensity, $I_f(\xi,\eta)=|E_f(\xi,\eta)|^2$, may be written as
\begin{equation}
    I_f(\xi,\eta) = 4 k^2\left| F_\text{infl}(\xi,\eta) \right|^2 \left|\sum_{j=1}^{N_\text{act}^\prime}\Delta V_j\;g_j\;e^{-ik(x_j\xi+y_j\eta)/f}\right|^2,
\end{equation}
\begin{equation}
    = 4 k^2\left| F_\text{infl}(\xi,\eta)\right|^2 \sum_{n=1}^{N_\text{act}^\prime}\sum_{m=1}^{N_\text{act}^\prime}c_{n,m} \cos\left(\frac{k}{f}\left(\Delta x_{n,m}\xi+\Delta y_{n,m}\eta\right)\right),
\end{equation}
where $F_\text{infl}(\xi,\eta)=\text{FT}\{f_\text{infl}(x,y)\}$, $c_{n,m}=g_n g_m \Delta V_n\Delta V_m$, $\Delta x_{n,m} = x_n-x_m$, and $\Delta y_{n,m} = y_n-y_m$. 
Assuming that the actuator gains are uniform (i.e. $g_n=g_m=g$) and that $\Delta V$ is an independent, identically distributed random variable:
\begin{equation}
    \sum_{n=1}^{N_\text{act}^\prime}\sum_{m=1}^{N_\text{act}^\prime}c_{n,m} \cos\left(\frac{k}{f}\left(\Delta x_{n,m}\xi+\Delta y_{n,m}\eta\right)\right) = N_\text{act}^\prime g^2 \sigma_{\Delta V}^2,
\end{equation}
where $\sigma_{\Delta V}^2$ is the variance of the voltage errors. Quantization errors are approximately uniformly distributed from $-V_\text{LSB}/2$ to $V_\text{LSB}/2$, where $V_\text{LSB}$ is the voltage corresponding to the least significant bit (LSB), and therefore have variance 
\begin{equation}
    \sigma_{\Delta V}^2=\frac{1}{V_\text{LSB}}\int_{-V_\text{LSB}/2}^{V_\text{LSB}/2} u^2 du = \frac{V_\text{LSB}^2}{12},
\end{equation}
where $u$ represents uniformly distributed $\Delta V$ values.

The relevant quantity for coronagraph performance is the raw contrast, which is often approximated by the so-called normalized intensity, $\hat{I}(\xi,\eta) = I(\xi,\eta)/I_0$, where for a circular Lyot stop,
\begin{equation}
    I_0 = \left( \frac{\pi b^2}{\lambda f} \right)^2 = \left( \frac{kb^2}{2f} \right)^2,
\end{equation}
and $b$ is the radius of the Lyot stop. The normalized intensity is therefore
\begin{equation}
    \hat{I}(\xi,\eta) = \frac{4 N_\text{act}^\prime f^2 h_\text{min}^2}{3 b^4} \left|F_\text{infl}(\xi,\eta)\right|^2,
\end{equation}
where $h_\text{min}=gV_\text{LSB}$ is the minimum DM surface motion.


\subsection{Influence function model \#1: simple Gaussian}

Assuming the influence function is a simple, peak-normalized Gaussian $f(x,y) = \exp(-(r/d)^2)$,
\begin{equation}
    \left| F_\text{infl}(\xi,\eta)\right|^2 = \frac{d^4 k^2}{4 f^2} \exp\left(-\frac{1}{2}\left(\frac{d k \rho}{ f}\right)^2\right),
    \label{eqn:FTinfl}
\end{equation}
where $\rho^2=\xi^2 + \eta^2$ and $d$ is the actuator radius. Near the optical axis, the exponential term is approximately one and the expected normalized intensity is
\begin{equation}
    \hat{I_0} = \frac{1}{3}N_\text{act}^\prime \frac{ d^4}{b^4}  k^2  h_\text{min}^2.
\end{equation}
We define the relative width of the influence function as $\omega=d/p$, where $p$ is the inter-actuator pitch, and $n_\text{act}$ as the number of influence functions across the Lyot stop. Substituting $b=n_\text{act}p/2$ and $N_\text{act}^\prime=\pi n_\text{act}^2/4$, the normalized intensity may then be written as
\begin{equation}
    \hat{I_0} = \frac{16\pi}{3n_\text{act}^2} \left( \frac{h_\text{min}}{\lambda} \right)^2  \pi^2\omega^4.
\end{equation}
The normalized intensity falls off as a function of angular separation from the optical axis, $\alpha$, with the profile of $\left| F_\text{infl}(\xi,\eta)\right|^2$, which is also known as the transfer function. By Eqn.~\ref{eqn:FTinfl}, $\hat{I} = \hat{I_0}\exp(-(\alpha/\alpha_\text{infl})^2)$, where
\begin{equation}
    \alpha_\text{infl} = \frac{\sqrt{2}}{2\pi}\frac{\lambda}{d} = \frac{\sqrt{2}}{2\pi}\frac{n_\text{act}}{\omega}\frac{\lambda}{D_\text{LS}} = \frac{\sqrt{2}}{2\pi}\frac{n_\text{act}}{\omega \Gamma}\frac{\lambda}{D},
\end{equation}
and $D_\text{LS}$ is the diameter of the Lyot stop. 

\subsection{Influence function model \#2: sum of two Gaussians}

Now, we model the influence function of the MEMS DM by a sum of two Gaussian functions:
\begin{equation}
    f(x,y) = c_1 \exp(-(r/d_1)^2) + c_2 \exp(-(r/d_2)^2),
\end{equation}
where $c_1$ and $c_2$ are constants and $d_1$ and $d_2$ are the radii of the Gaussian functions. Following a similar derivation as the previous case, 
\begin{equation}
    \hat{I} = \frac{16\pi}{3n_\text{act}^2} \left( \frac{h_\text{min}}{\lambda} \right)^2 \pi^2 \Phi(\alpha),
    \label{eqn:Itwogaussians}
\end{equation}
where 
\begin{equation}
    \Phi(\alpha) =  c_1^2 \omega_1^4 e^{-(\alpha/\alpha_1)^2} + c_2^2 \omega_2^4 e^{-(\alpha/\alpha_2)^2} + 2c_1c_2 \omega_1^2\omega_2^2 e^{-(\alpha/\alpha_3)^2},
    \label{eqn:Phi}
\end{equation}
\begin{equation}
    \omega_1 = d_1/p,
\end{equation}
\begin{equation}
    \omega_2 = d_2/p,
\end{equation}
\begin{equation}
    \alpha_1 = \frac{\sqrt{2}}{2\pi}\frac{\lambda}{d_1} = \frac{\sqrt{2}}{2\pi}\frac{n_\text{act}}{\omega_1 \Gamma}\frac{\lambda}{D},
\end{equation}
\begin{equation}
    \alpha_2 = \frac{\sqrt{2}}{2\pi}\frac{\lambda}{d_2} = \frac{\sqrt{2}}{2\pi}\frac{n_\text{act}}{\omega_2 \Gamma}\frac{\lambda}{D},
\end{equation}
\begin{equation}
    \alpha_3 = \frac{1}{\pi}\frac{\lambda}{\sqrt{d_1^2+d_2^2}} = \frac{1}{\pi \Gamma}\frac{n_\text{act}}{\sqrt{\omega_1^2 + \omega_2^2}}\frac{\lambda}{D}.
\end{equation}



\acknowledgments     
This work was carried out at the Jet Propulsion Laboratory, California Institute of Technology, under contract with the National Aeronautics and Space Administration (NASA).


\bibliography{refLibrary}   
\bibliographystyle{spiebib}   

\end{document}